\documentstyle[multicol,graphicx,aps]{revtex}

\begin{document}
\draft
\title{Ordering of the three-dimensional Heisenberg
      spin glass in magnetic fields}
\author{Hikaru Kawamura\footnote{E-mail : kawamura@ess.sci.osaka-u.ac.jp}
 and Daisuke Imagawa}
\address{Department of Earth and Space Science, Faculty of Science,
Osaka University, Toyonaka 560-0031,
Japan}
\date{\today}
\maketitle
\begin{abstract}
Spin and chirality orderings of the three-dimensional Heisenberg
spin glass are studied under magnetic fields
in light of the recently developed
spin-chirality decoupling-recoupling scenario.
It is found by Monte Carlo
simulations that the chiral-glass transition
and the chiral-glass ordered state,
which are essentially of the same character
as their zero-field counterparts,
occur under magnetic fields.
Implication to experimental phase diagram is discussed.
\end{abstract}
\begin{multicols}{2}
\narrowtext

In the studies of spin glasses, much effort has been devoted
to the properties  under magnetic fields.
Unfortunately, our understanding of them
still has remained unsatisfactory\cite{Review}.
Most of the numerical studies
have  focused on the properties of the
simple Ising model,
especially the three-dimensional (3D) Edwards-Anderson (EA) model.
While the existence of a true thermodynamic spin-glass (SG) transition
has been established  for this model in zero field,
the question of its existence or nonexistence in magnetic fields
has remained unsettled.

If one tries to understand real
experimental SG ordering,
one has to remember that many of real SG materials are
more or less Heisenberg-like rather than Ising, in the sense that
the random magnetic anisotropy is considerably weaker than
the isotropic exchange interaction\cite{Review,OYS}.
Numerical simulations
have indicated that the
isotropic 3D Heisenberg SG with finite-range interaction
does not exhibit the conventional
SG order at finite
temperature in zero field\cite{Review,OYS,Kawamura92,Kawamura98,HK1}.
Since applied fields
generally tend to suppress the SG ordering,
a true thermodynamic SG transition is even more unlikely
under magnetic fields in case of the Heisenberg model.

Experimentally, however, a rather sharp transition-like behavior
has been observed under magnetic fields
in typical Heisenberg-like SG magnets, {\it e.g.\/},
canonical SG like AuFe and CuMn,
although it is not completely clear whether
the observed anomaly corresponds
to a true thermodynamic transition\cite{Review,Orbach,Campbell}.
The situation is in contrast to  the zero-field case
where the existence of a true thermodynamic SG transition has
been established experimentally\cite{Review}.
Set aside the question of 
the strict nature of the SG ``transition'', it is experimentally
observed that  a weak applied field
lowers the zero-field SG transition temperature rather quickly
\cite{Review,Orbach,Campbell}.
For higher fields, the SG ``transition''
becomes much more robust to fields,
where the ``transition temperature'' shows much less field dependence
\cite{Review,Orbach,Campbell}.
Such  behaviors of the SG transition temperature under magnetic fields
$T_g(H)$ were often interpreted in terms of the mean-field model
\cite{Review,Orbach}. Indeed,
the mean-field Sherrington-Kirkpatrick (SK) model with
an infinite-range Heisenberg exchange interaction with weak
random magnetic anisotropy
exhibits a transition line
similar to the experimental one \cite{Kotliar}:
{\it i.e.\/}, the so-called de Almeida-Thouless (AT) line
$H\propto (T_g(0)-T_g(H))^{3/2}$
in weak-field regime where the anisotropy is important,
and the Gabay-Toulouse (GT) line
$H\propto (T_g(0)-T_g(H))^{1/2}$ in strong-field regime
where the anisotropy is unimportant.
Nevertheless, if one notes that the true finite-temperature transition under
magnetic fields, though
possible in the infinite-range SK model, 
is unlikely to occur
in a more realistic finite-range Heisenberg model,
an apparent success of the mean-field model in explaining the
experimental phase diagram should be taken with strong reservation.

Recently, one of the present authors has proposed a scenario,
the spin-chirality decoupling-recoupling scenario, aimed at explaining
some of the puzzles
concerning the experimentally observed SG transition in zero field
\cite{Kawamura92}.
In this scenario, {\it chirality\/}, which is a multispin variable
representing  the sense or the handedness of local
noncoplanar spin structures induced by spin frustration,
plays an essential role.
In a fully isotropic Heisenberg SG, in particular,
this scenario claims the occurrence of a novel {\it chiral-glass\/}
ordered state in which only the chirality exhibits a glassy long-range
order (LRO) while the spin remains paramagnetic. 
At the chiral-glass transition,
among the global symmetries of the Hamiltonian,
only the $Z_2$ spin reflection (inversion)
symmetry is broken spontaneously
with keeping the $SO(3)$ spin rotation symmetry preserved.
Note that this picture
entails the spin-chirality (or $SO(3)-Z_2$)
decoupling on long length and time scales:
Namely, although the chirality is not independent of the spin
on microscopic length scale,
it eventually exhibits a long-distance behavior
entirely different from the spin.
Such a chiral-glass transition
was indeed observed in zero field
in a recent Monte Carlo simulation
by Hukushima and Kawamura
\cite{HK1}. It was also found there
that the critical properties associated with the chiral-glass transition
were different from those of the Ising SG, and that the chiral-glass
ordered state exhibited a one-step-like novel RSB.

In the chirality scenario of Ref.\cite{Kawamura92},
experimental SG
transition in real Heisneberg-like SG magnets is regarded essentially as
a chiral-glass transition 
``revealed'' via the magnetic anisotropy.
Weak but finite random magnetic anisotropy inherent to real 
magnets
``recouples'' the spin to the chirality,
and the chiral-glass transition shows up
as an experimentally observable
{\it spin\/}-glass transition.

The purpose of the present Letter is to
reexamine the SG ordering of the 3D isotropic Heisenberg SG under
applied fields in light of the above chirality scenario.
We first argue some of the possible consequences of the
chirality picture on the finite-field SG properties,
and then perform extensive Monte Carlo (MC) simulations
to check how the spin and chirality really
order in fields.
We show that the chiral-glass  phase, essentially of
the same character as the zero-field one, remains stable as
a true thermodynamic phase in
applied fields. It is also found that
the associated chiral-glass transition line possesses some of
the character
of the GT line of the mean-field model, yet its  physical origin
entirely different.

Though we expect that our argument  holds quite generally,
we fix here our model Hamiltonian, which is just the one
used in our
MC simulation  below.
We consider the isotropic classical Heisenberg
model on a 3D simple cubic lattice,
\begin{equation}
{\cal H}=-\sum_{<ij>}J_{ij}\vec S_i\cdot \vec S_j - H\sum_iS_i^z,
\end{equation}
where $H$ is the intensity of the magnetic field applied along
the $z$ direction. 
The  nearest-neighbor coupling $J_{ij}$ is assumed to
take the value $J$ or $-J$ with equal probability ($\pm J$ distribution).
Local scalar chirality is defined for three neighboring spins:
Here we define it
at the $i$-th site and in the $\mu $-th direction ($\mu =x,y,z$)
by $\chi _{i\mu }=\vec
S_{i-\hat e_\mu}\cdot \vec S_i\times \vec S_{i+\hat e_\mu}$,
$\hat e_\mu $ being a unit lattice vector in the $\mu $-th direction.

Applied fields reduce
the global symmetry of the Hamiltonian (1)
from the zero-field one $Z_2\times SO(3)$
to $Z_2\times SO(2)$, where
$Z_2$ refers to the spin reflection
with respect to an arbitrary plane including the field axis, and
$SO(2)$ to the spin rotation around the field axis.
Note that, even in fields,
the chiral $Z_2$ reflection symmetry, characterized
by the mutually opposite signs of
the chirality, is kept intact. The chirality is a pseudoscalar
invariant under  $SO(2)$ (or $SO(3)$ in case of
$H=0$) rotations but changes sign under $Z_2$ reflections.
The same  chiral $Z_2\times SO(2)$ symmetry also
appears in the $XY$ SG with two-component  spins\cite{XYCG}.


Since the chiral $Z_2$ is supposed to be decoupled from the spin $SO(3)$
already in zero field, and
the applied field serves only to reduce
the decoupled $SO(3)$ to $SO(2)$, one naturally
expects that the $Z_2$ chiral-glass transition,
essentially of the same type as
the zero-field one, would persist even under magnetic fields
\cite{Comment}.
More specifically,
the chiral-glass transition in a field should lie in the same
universality class as the zero-field one, characterized
by the same set of exponents, and the chiral-glass
ordered state in a field exhibits the same kind of one-step-like
RSB as in zero field.

Since the chiral-glass order in  a field is expected to be
essentially of the same character as in  zero field,
the chiral-glass transition
temperature under magnetic fields $T_{{\rm CG}}(H)$
should be a regular function of $H$.
If one takes account of the obvious symmetry
$H\leftrightarrow -H$, the chiral-glass transition line at low
enough fields should behave as
\begin{equation}
T_{{\rm CG}}(0)-T_{{\rm CG}}(H)=cH^2+c'H^4+\cdots .
\end{equation}
Generally, the coefficient $c$ could be either positive or negative.
Interestingly, the above form of the
chiral transition line is  similar to the  so-called
GT line of the mean-field Heisenberg SK model.
We emphasize, however, that
their physical origin is entirely different.
The quadratic dependence of the chiral-glass transition line
is simply of regular origin, whereas that of the GT-line in the
SK model cannot be regarded so.

Concerning the spin order, applied
fields trivially induce a nonzero
longitudinal (parallel to the field) spin order at any
temperature. The behavior of the transverse  (perpendicular to the field)
component
could be more nontrivial.
If one recalls the recent numerical results
on the 3D  {\it XY\/} SG
which indicate the absence of the standard SG order
at least just below the chiral-glass transition \cite{XYCG,Maucourt},
the transverse
spin order is also unlikely to arise in the present case,
at least just below the $Z_2$ chiral-glass transition.

In order to examine whether the above expectation
really holds or not,
we next perform extensive MC simulations on the isotropic
$\pm J$ Heisenberg SG model (1).
Simulations are performed for a variety of fields
$H/J=0.05, 0.1, 0.5, 2.0, 3.0, 5.0$.
The lattices studied are simple-cubic lattices with $L^3$ sites
with $L=6\sim 16$ with periodic boundary conditions.
The system is fully equilibrated with
use of the temperature-exchange method\cite{HN}.
We show here the data explicitly
for the particular field value of $H/J=0.5$, where the
sample average is taken over 128-800
bond realizations.

By running two independent sequences of  systems
(replica 1 and 2) in parallel, we
compute a scalar chiral overlap $q_\chi$
between the chiralities of the two replicas by
$q_{\chi} = \frac{1}{3N}\sum_{i\mu}\chi_{i\mu}^{(1)}\chi_{i\mu}^{(2)}$,
as well as a  spin-overlap tensor $q_{\mu\nu}$
between the transverse $\mu$ and $\nu$
components  ($\mu$, $\nu$=$x,y$) of the spin by
$q_{\mu\nu}=\frac{1}{N}\sum_i S_{i\mu}^{(1)}S_{i\nu}^{(2)}$.
Then, in terms of these overlaps,
we calculate the  Binder ratios of the chirality $g_\chi $, and
of the transverse ($XY$) components of the
spin $g_T$ defined in the standard manner: See Ref.\cite{HK1} for detailed
definition. The results are shown in Fig.1.
The Binder ratio of the chirality $g_\chi $ 
exhibits a negative dip which, with increasing $L$,
tends to deepen and shift toward lower
temperature.  Furthermore, $g_\chi $
of various $L$ cross at a temperature slightly above the
dip temperature
$T_{{\rm dip}}$  {\it on negative side of
$g_\chi$\/},  eventually merging at temperatures lower than $T_{{\rm dip}}$.
The observed  behavior of $g_\chi $ is similar to the one
observed in zero field\cite{HK1}. As argued in
Ref.\cite{HK1}, the persistence of a negative dip and the crossing occurring
at $g_\chi<0$, are strongly
suggestive of the occurrence of a finite-temperature transition
where $g_\chi (T_{{\rm CG}}^-)$ and $g_\chi (T_{{\rm CG}})$
take {\it negative\/} values in the $L\rightarrow \infty $ limit.
In the inset of Fig.1(a), we plot the  negative-dip
temperature $T_{{\rm dip}}(L)$
versus $1/L$. The data lie on a
straight line fairly well, and its extrapolation to $1/L=0$
gives an estimate of the bulk chiral-glass
transition temperature, $T_{{\rm CG}}/J\sim 0.25$.
(More precisely, $T_{{\rm dip}}(L)$ should scale
with $L^{1/\nu }$ where $\nu $ is the chiral-glass correlation-length
exponent. As shown below, our estimate of $\nu \simeq 1.3$
comes close to unity, more or less
justifying the linear extrapolation employed here. Extrapolation with respect
to $L^{1/1.3}$ yields $T_{{\rm CG}}/J\sim 0.23$.)

In sharp contrast to $g_\chi $, Binder ratio of the
transverse component of the spin
$g_T$ decreases monotonically toward zero with increasing $L$,
without a negative dip nor a crossing,
suggesting that the transverse component of spin remains
disordered even below $T_{{\rm CG}}$.

\begin{center}
\begin{figure}[t]
\includegraphics[width=9cm,height=10cm]{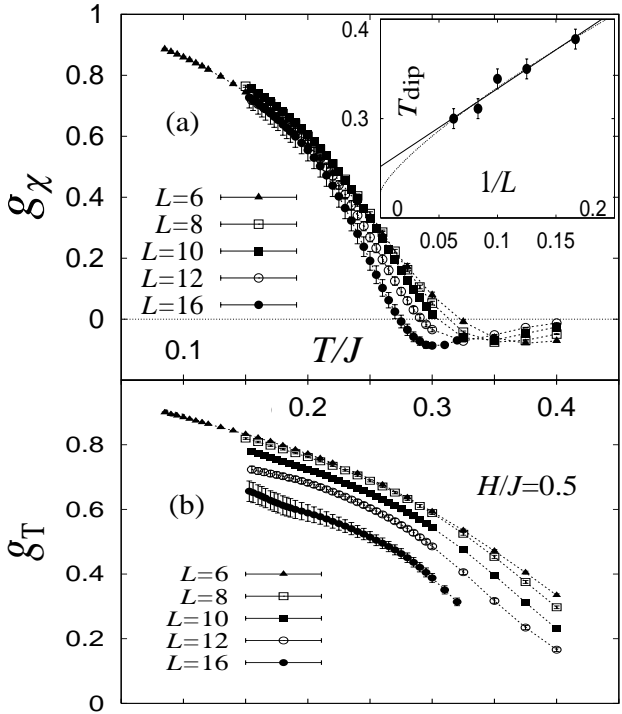}
\caption{
The temperature and size dependence
of the Binder ratios of the chirality (a), and of the
transverse ($XY$) component of the spin (b), in
a field 
$H/J=0.5$. Inset displays the negative-dip temperature
vs. $1/L$. The solid and broken lines are the best fits
assuming the $1/L$ and $1/L^{1/1.3}$ dependence, respectively.
}
\end{figure}
\end{center}


Validity of our estiamte of
$T_{{\rm CG}}$ has been checked also from the behavior of
other quantities,
{\it e.g.\/}, the temporal-decay of the equilibrium
chirality autocorrelation function (data not shown here),
which yields $T_{{\rm CG}}/J=0.23(2)$ in accord with the above
estimate.
In order to probe the possible RSB in the chiral-glass ordered state,
we display in Fig.2 the distribution function of the chiral-overlap
defined by
$P(q'_\chi )=[\langle\delta (q_\chi -q'_\chi )\rangle]$
where $<\cdots>$  and [$\cdots$] represent the thermal average and
the sample average, respectively.
The existence of a growing ``central peak'' at $q_\chi =0$ for larger $L$,
in addition to the standard ``side-peaks''
corresponding to $\pm q_{{\rm CG}}^{{\rm EA}}$,
suggests the occurrence of a one-step-like peculiar RSB in the chiral-glass
ordered state. Similar behavior was
observed in the chiral-glass state
in zero field\cite{HK1}.
The existence of a persistent negative
dip in the Binder ratio $g_\chi $
is also
consistent with the occurrence of such a one-step-like RSB\cite{HK2}.

\begin{center}
\begin{figure}[h]
\includegraphics[width=9cm,height=7cm]{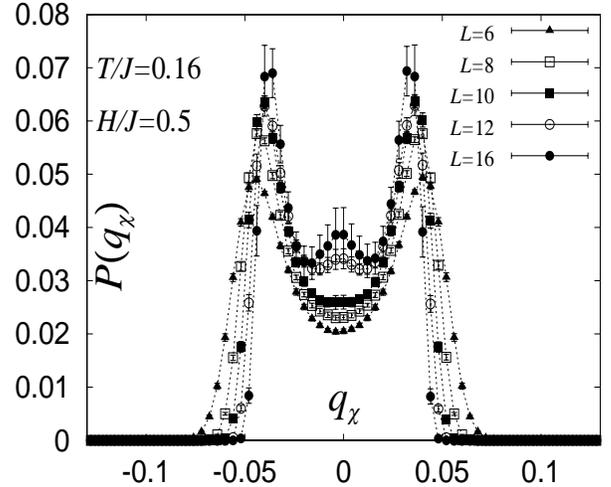}
\caption{
Chiral-overlap distribution function at $T/J=0.16$,
well below the chiral-glass transition point
$T_{{\rm CG}}/J\simeq 0.23$, in a field
of $H/J=0.5$.
}
\end{figure}
\end{center}
With setting $T_{{\rm CG}}/J=0.23$ as determined above,
we perform the standard finite-size scaling 
of the chiral-glass order parameter $[<q_\chi^2>]$ and of the chiral
autocorrelation function, to estimate various
chiral-glass exponents. We then
get $\nu= 1.3(2)$, $\eta=0.6(3)$, $z=5.3(5)$, which turns out to
agree within errors with the corresonding
zero-field exponents of Ref.~\cite{HK1}.
The results seem consistent
with a  common universality class occurring both
in zero-field and finite-field chiral-glass transitions.

Similar calculations and analysis are repeated for other field
values as well. For $H/J=0.1$, in particular,
we have performed the same scale of intensive calculation as
was done for $H/J=0.5$, to find that all
qualitative features are similar.
The chiral-glass transition with a one-step-like RSB occurs
at $T_{{\rm CG}}/J=0.21(2)$,
with the exponents
$\nu= 1.3(2)$, $\eta= 0.6(3)$, $z= 4.9(5)$,
which agrees within errors
with our estimates for  $H/J=0.5$.

By collecting the $T_{{\rm CG}}$ values determined
for other field values,
we construct a phase
diagram in the temperature-magnetic field plane as shown
in Fig.3.
The chiral-glass state
remains quite robust against magnetic fields. Indeed,
$T_{{\rm CG}}(H)$ is not much reduced from the zero-field value
even at field as large as ten times of $T_{{\rm CG}}(0)$.
This somewhat surprising property probably
arises from the fact that
the magnetic field
couples in the Hamiltonian directly to the spin,
{\it not to the chirality\/},
and the effective coupling between the field and the chirality is rather weak.
At lower fields, the chiral-glass
transition line is almost orthogonal to the $H=0$ axis,
consistent with the expected behavior Eq.(2). Our data are
even not inconsistent with the coefficient $c$ in Eq.(2)
being slightly
negative so that
$T_{{\rm CG}}(H)$ initially {\it increases\/} slightly with $H$,
though it is difficult to draw a
definite conclusion due to the scatter
of our estimate of $T_{{\rm CG}}(H)$.

\begin{center}
\begin{figure}[h]
\includegraphics[width=9cm,height=6cm]{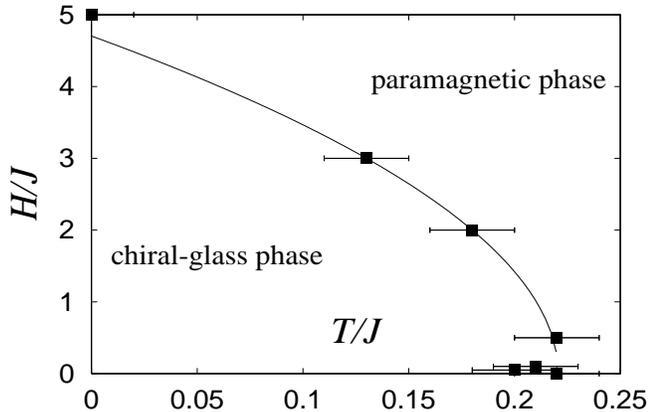}
\caption{
The temperature - magnetic field
phase diagram of the 3D isotropic $\pm J$
Heisenberg spin glass.
Note the difference in
energy scales of the temperature- and the field-axes.}
\end{figure}
\end{center}

Finally, we wish to discuss  implications of our
results 
to real experimental SG.
In real Heisenberg-like SG magnets,  weak but finite random magnetic
anisotropy neglected here
plays a role.
Among other things,
the anisotropy recouples the spin to the chirality,
transforming the chiral-glass state
into the {\it spin\/}-glass state. Furthermore,
in the presence of both random magnetic anisotropy and
magnetic field, all global symmetries of the Hamiltonian
will be lost.
Nevertheless, we expect that
the chiral-glass transition should still
persist in fields as a pure RSB transition, not accompanying
the global $Z_2$-symmetry breaking.
Difference in the broken symmetries in zero- and in finite
fields, however,
causes a singular crossover behavior in fields,
in sharp contrast to the fully isotropic case.
We believe that this singular crossover line expected in the
weak-field regime of anisotropic system
is nothing but the AT-like transition line ubitiously observed
experimentally.
Further detailed nature of this crossover expected in the low-field regime
will be discussed in a separate paper\cite{Kawamura00}.

Meanwhile, in the strong-field
regime where the applied field dominates
the anisotropy, main features of
the experimental phase diagram would be described by our present results
on isotropic system. Thus, the experimental observation of the GT-like
field-insensitive
transition line in the high-field regime is fully consistent with our
observation of the
field-insensitive chiral-glass transition line\cite{Review,Orbach,Campbell}.
Other interesting possibility revealed by our analysis is
that the SG transition line might extend to higher fields
than hitherto suspected. For example, for canonical SG AuFe,
Campbell {\it et al\/} recently
determined the SG phase boundary in  fields
by torque measurements\cite{Campbell}:
At lower fields,  applied fields
rapidly suppress the SG transition giving rise to the standard AT-like
behavior, whereas around the maximal fields of the measurements ($\sim 7$T)
the SG transition line becomes
almost-field independent. If one roughly estimates the effective $J$
of this system from its
zero-field SG transition temperature, it is of order 50K.
If one assumes that the energy scale of our present model calculation
could roughly be applicable to AuFe, the SG ordered state should
extend to fields much higher than $10$T
without much reduction, or even
with a slight increase, in $T_g$,
although, considering the difference in microscopic
details between the present model and real AuFe, one cannot
expect a truely quantitative correspondence.
Anyway, further high-field experiments on AuFe
and on other Heisenberg-like SG magnets might be worthwhile
to determine the
SG phase boundary in the high-field regime.

The numerical calculation was performed on the Hitachi SR8000 at the
supercomputer center, ISSP, University of Tokyo. The authors are
thankful to Dr.K. Hukushima for useful discussion.

\end{multicols}

\begin{thebibliography}{99}
\bibitem{Review}
For reviews on spin glasses, see {\it e. g., }
(a) K. Binder and A. P. Young, Rev. Mod. Phys. {\bf 58}, 801 (1986);
(b) K. H. Fischer and J. A. Hertz, {\it Spin Glasses} Cambridge
University
Press (1991);
(c) J. A. Mydosh, {\it Spin Glasses} Taylor \& Francis (1993);
(d) A. P. Young ({\it ed.\/}),
{\it Spin glasses and random fields} World Scientific,
Singapore (1997).


\bibitem{OYS}
J. A. Olive, A. P. Young and D. Sherrington, Phys. Rev. B{\bf 34},
6341 (1986).




\bibitem{Kawamura92}
H. Kawamura, Phys. Rev. Lett.  {\bf 68}, 3785 (1992);
Int. Jour. Mod. Phys. {\bf 7}, 345 (1996).


%
%
%


%






\bibitem{Kawamura98}
H. Kawamura, Phys. Rev. Lett. {\bf 80}, 5421 (1998).


\bibitem{HK1}
K. Hukushima and H. Kawamura, Phys. Rev. E{\bf 61}, R1008 (2000).


\bibitem{Orbach}
D. Chu, G.G. Kennig and R. Orbach, Phys. Rev. Lett.
{\bf 72}, 3270 (1994).


\bibitem{Campbell}
D. Petit, L. Fruchter and I.A. Campbell, Phys. Rev. Lett.
{\bf 83}, 5130 (1999).


\bibitem{Kotliar}
G. Kotliar and H. Sompolinsky,, Phys. Rev. Lett
{\bf 53}, 1751 (1984).


\bibitem{XYCG}
H. Kawamura, Phys. Rev. {\bf B} 51, 12398 (1995);
H. Kawamura and M.S. Li, unpublished.

\bibitem{Comment}
One can consider here
a more complicated possibility
that the change of the symmetry $SO(3)\rightarrow SO(2)$
affects the $Z_2$ chiral part in a nontrivial way, say,
seriously modifies the nature of the chirality-chirality interaction,
and might lead to the finite-field chiral transition qualitatively
different from the zero-field one.
However, we restrict ourselves here
to the simplest possibility neglecting such possible complications.

\bibitem{Maucourt}
J. Maucourt and D. R. Grempel, Phys. Rev. Lett. {\bf 80}, 770 (1998).

\bibitem{HK2}
K. Hukushima and H. Kawamura, Phys. Rev. E{\bf 62}, 3360 (2000).





%


\bibitem{HN}
K. Hukushima and K. Nemoto, J. Phys. Soc. Jpn. {\bf 65}, 1604 (1996).


\bibitem{Kawamura00}
H. Kawamura, unpublished.







\end{thebibliography}
\end{document}